\documentclass[aps,pre,twocolumn,groupedaddress]{revtex4-1}
\usepackage[english]{babel}
\usepackage{setspace}
\usepackage{graphicx}
\usepackage{epstopdf}
\usepackage{color}

\begin{document}
\title{High-Brightness Megahertz-Rate Beam from a Direct-Current and Superconducting Radio-Frequency Combined Photocathode Gun}
\author{H. Jia}
\author{T. Li}
\author{T. Wang}
\author{Y. Zhao}
\author{X. Zhang}
\author{H. Xu}
\author{Z. Liu}
\author{J. Liu}
\author{L. Lin}
\author{H. Xie}
\author{L. Feng}
\author{F. Wang}
\author{F. Zhu}
\author{J. Hao}
\author{S. Quan}
\author{K. Liu}
\author{S. Huang}\email{huangsl@pku.edu.cn}
\affiliation{State Key Laboratory of Nuclear Physics and Technology and Institute of Heavy Ion Physics, School of Physics, Peking University, Beijing 100871, China}

\begin{abstract}
High-brightness megahertz-rate electron beams are highly desired for cutting-edge applications in many areas of science. Photocathode electron guns capable of generating such beams with low dark current remain to be a challenging field. In this Letter we report the breakthroughs with a hybrid gun combining a direct-current gap and a superconducting radio-frequency (SRF) cavity. The gun, employing K${}_2$CsSb photocathodes driven by a green laser, delivers a few MeV electron beam at 1 MHz and 81.25 MHz rates with an average current up to 3 mA and a dark current several orders lower than current normal conducting continuous-wave guns. Emittance compensation and mutipole field corrections have been applied to achieve a high-quality beam. The achievement will inspire the development of high-brightness SRF guns and promote megahertz-rate beam applications including new generation of coherent linac light source and ultrafast electron diffraction/microscopy. 
\end{abstract}

\maketitle

Photocathode guns are the most important high brightness electron sources~\cite{rao2014engineering, Stephan2014}, whose technical advancement has greatly promoted the development and applications of electron accelerators, such as {x-ray free-electron laser (XFEL)}~\cite{LCLS2010, RevModPhys.88.015007, EXFEL2020, SWISSFEL2020}, energy recovery linac (ERL)~\cite{NEIL20069, AKEMOTO2018197}, ultrafast electron diffraction~\cite{RevModPhys.94.045004}, etc. The achievements made in these applications in turn have placed higher demands on photocathode guns~\cite{10.1063/1.4789395, 10.3389/fphy.2023.1150809, 10.3389/fphy.2023.1166179}. 
Especially, generating megahertz (MHz)-rate high brightness electron beams with low dark current becomes a hot topic during the past two decades~\cite{PhysRevLett.102.104801, 10.3389/fphy.2023.1150809, PhysRevAccelBeams.24.124402, 10.3389/fphy.2023.1166179, PhysRevAccelBeams.24.033401, PhysRevLett.124.244801}.

To accelerate MHz-rate beams, continuous-wave (CW) operation of a radio-frequency (RF) cavity is needed, which adds great challenges for a normal conducting (NC) RF gun due to huge heat load and high dark current~\cite{10.3389/fphy.2023.1150809}. 
On the other hand, superconducting RF (SRF) guns are the natural candidate for CW RF operations due to the low power dissipation on the cavities~\cite{PhysRevLett.124.244801, CHALOUPKA1989327, CALAGA2006159, PhysRevSTAB.14.053501, PhysRevAccelBeams.24.033401}. However, outstanding challenges still remain, such as photocathode integration into the SRF structure, emittance compensation, etc.~\cite{10.3389/fphy.2023.1166179}

In this Article, we report the latest breakthroughs achieved with a direct-current (DC) and SRF combined photocathode gun at Peking University. We will first present a brief overview of the design considerations and main features. Then we will show the commissioning results. Especially, we demonstrate the emittance compensation~\cite{CARLSTEN1989313, PhysRevE.55.7565, PhysRevLett.99.234801} and multipole magnetic field corrections~\cite{PhysRevAccelBeams.21.010101,PhysRevAccelBeams.25.084001}, which are crucial for achieving low beam emittance as required by CW {XFELs}. 

\begin{figure*}[htbp]
\begin{centering}
\includegraphics[width=1.0\textwidth]{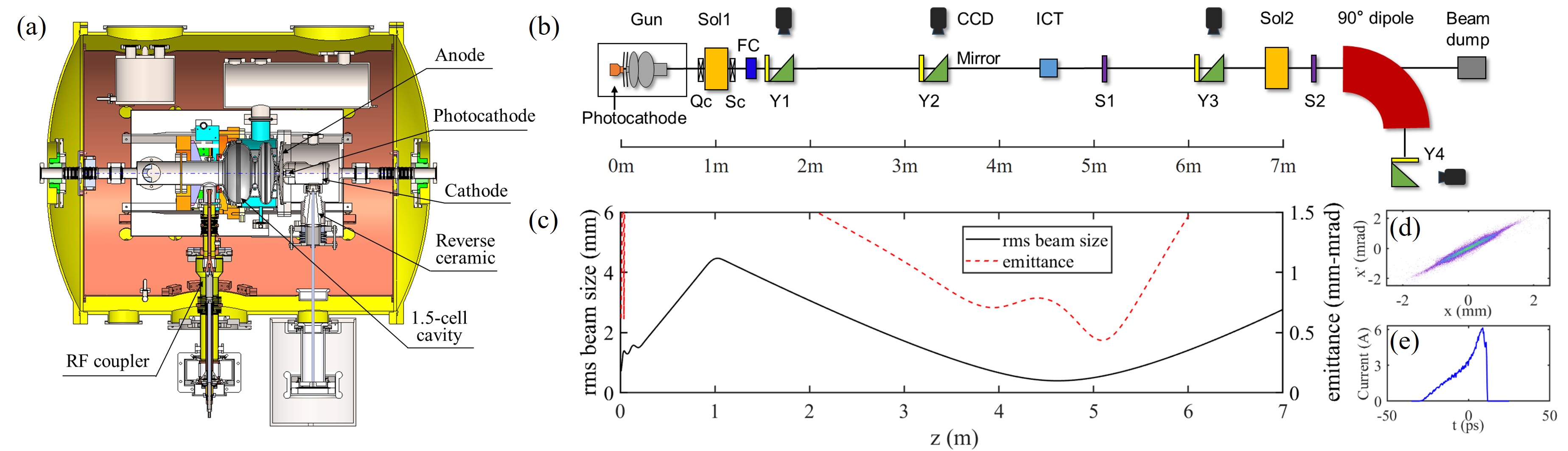}\\
\end{centering}
\vspace{-4mm}
\caption{\label{fig:sketch} A sketch of the DC-SRF-II gun (a), the electron beam line (b), the simulated beam size and emittance along the beam line (c), and the {simulated} phase space (d) and current profile (e) at S1. The beam line comprises two solenoid lenses (Sol1-Sol2), quadrupole coils (Qc), sextupole coils (Sc), a 90${}^\circ$ dipole magnet, four pneumatic YAG screens (Y1-Y4), two motorized molybdenum plate with single slit (S1-S2), a Faraday cup (FC), an integrated current transformer (ICT), and a beam dump.}
\vspace{-5mm}
\end{figure*}

The concept of DC-SRF gun was originally proposed in 2001 to address the problem of compatibility between semiconductor photocathodes and SRF cavity~\cite{ZHAO2001564}. It combines a pair of DC high voltage electrodes and a 1.3 GHz SRF cavity connected by a short drift tube (about 10 mm long), as shown in Fig.~\ref{fig:sketch}(a). The photocathode is located in the DC gap and therefore separated from the cavity. This design has a few advantages: 1) it allows to use a normal conducting photocathode in an SRF gun, and isolation with an RF choke filter~\cite{10.3389/fphy.2023.1166179} is not needed any more; 2) it avoids the potential contamination of SRF cavity from the semiconductor materials; 3) it greatly diminishes the dark current arising from the rim of photocathode plug or the cathode nose. The short distance between the DC gap and the cavity allows the required DC voltage to be a few tens to 100 kV, which puts less stringent requirements on high voltage components and makes the gun very compact. 
{Note that although the voltage is lower than DC photocathode guns, the cathode field is similar or even higher due to the smaller gap and the electron energy can reach a few MeV at the exit of the gun.}
Besides, the DC-SRF hybrid structure provides an excellent vacuum environment for sensitive photocathodes, especially bi-alkali photocathodes which have significant quantum efficiency (QE) in green region of light spectrum. This helps greatly reduce the requirements on drive lasers for MHz-rate operation. 

The development of DC-SRF gun has undergone three stages: the prototype for feasibility test~\cite{XIANG2004321, HAO2006138}, the first generation (DC-SRF-I) achieving stable operation in pulse mode~\cite{QUAN2015117}, and the second generation (DC-SRF-II) attaining low emittance CW operation. Here the DC-SRF-II gun represents a milestone. The operating voltage of the DC gap has been raised to 100 kV and the electric field at the photocathode surface is 6 MV/m accordingly. The SRF cavity, which has 1.5 cells, {can be operated at an on-axis peak electric field ($E_\textrm{max}$) above 22 MV/m in CW mode.} For comparison, the DC-SRF-I gun could only be operated with a maximum DC voltage of 50 kV and {a highest SRF cavity gradient of 9 MV/m~\cite{HUANG2022}, which corresponds to an $E_\textrm{max}$ of about 15 MV/m.} These significant advancements lay an important foundation for achieving low emittance beams.

As a substantial change to the previous version, the DC-SRF-II gun adopts K${}_2$CsSb as photocathode material instead of Cs${}_2$Te. The K${}_2$CsSb photocathodes prepared for the gun have a QE above 5\% at 515 nm~\cite{OUYANG2022166204}, providing tremendous flexibility in the design of drive laser. For the DC-SRF-II gun, a drive laser based on an all-fiber master oscillator power amplifier has been developed~\cite{Wang:24}, which can deliver CW laser pulses at 1 MHz or 81.25 MHz rate and provide pulse trains with flexible timing patterns for gun commissioning. More importantly, the drive laser can reliably operate at an average power up to 10 W. The sufficient margin for power loss allows the utilization of high-quality laser shaping, which is of essential importance for optimizing the low emittance electron beam. Besides, it is worth noting the K${}_2$CsSb photocathodes are expected to have a lower intrinsic emittance compared to Cs${}_2$Te photocathodes as widely used in normal conducting RF guns, which would also help reduce the beam emittance~\cite{XIANG201558, Wang2020}.

The commissioning of the gun was started in 2021. As one of the major concerns for CW operation, the dark current from the gun, originating from the field emission on the inner surface of the DC gap and SRF cavity, has been carefully investigated. Experiments were first performed to evaluate the dark current from the DC gap under the same condition of CW operation while the drive laser was purposely blocked. In the measurement, the DC voltage was at 100 kV, while the cavity was operated with a lower {$E_\textrm{max}$ close to 15 MV/m} so as to mitigate its contribution. The field-emitted electrons, focused by a solenoid lens at the exit of the gun (Sol1), were collected by a Faraday cup (FC) as shown in Fig.~\ref{fig:sketch}(b), and the current was recorded by a picoammeter with a resolution of 0.1 pA. The solenoid strength was carefully scanned over a large range, while the readout of the picoammeter remained at zero, indicating the dark current from the DC gap was less than 0.1 pA. 

\begin{figure}[htbp]
\vspace{-3mm}
\begin{centering}
\includegraphics[width=0.4\textwidth]{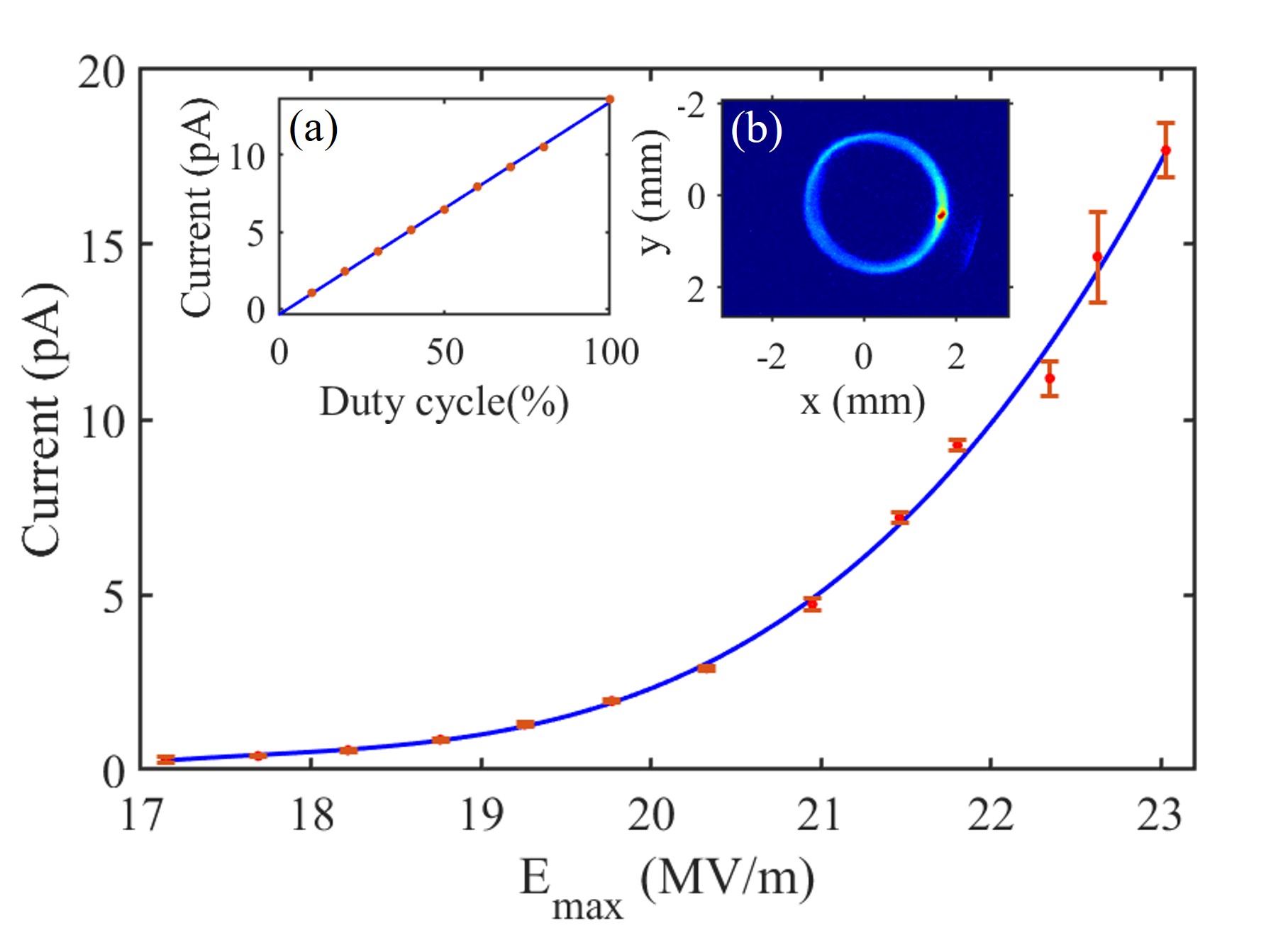}\\
\par\end{centering}
\vspace{-4mm}
\caption{\label{fig:darkcurrent} {Measured dark current} vs cavity gradient in pulsed mode with a duty factor of 10\%. Inset (a) shows the dark current vs RF duty factor {at the $E_\textrm{max}$ of 19 MV/m}; (b) shows an image of the field-emitted electrons.}
\vspace{-3mm}
\end{figure}

The dark current investigation was then focused on the SRF cavity only. Fig.~\ref{fig:darkcurrent} shows a typical measurement of dark current as a function of the cavity gradient, where the DC voltage was zeroed and the cavity was operated in pulsed mode with a duty factor of 10\% {for safety considerations of RF system}. The measurement results can be extrapolated to CW mode, since the dark current has a linear dependency on the RF duty factor (see Inset (a)). It can be inferred that the dark current is below 100 pA when the cavity is operated in CW mode with {an $E_\textrm{max}$ up to 22 MV/m}. An image of the field-emitted electrons is also shown in Fig.~\ref{fig:darkcurrent}, for which the electrons were focused onto a YAG screen (Y1) right after the Faraday cup. The solenoid strength was close to that required to focus the photoelectrons from the gun, implying a similar electron energy, while the ring-shaped profile further suggests the electrons were emitted around the entrance iris of the cavity. It is worth noting there is a stronger emission region in the lower right of the dark current ring, indicating a defect area therein. However, this also means the dark current might be reduced through a careful processing of the cavity. Note this level of dark current ($\sim$100 pA) is about 4 orders lower than that in the current normal conducting CW guns such as the one at the LCLS-II (a few ${\mu}$A level)~\cite{PhysRevAccelBeams.24.073401}. We believe such a low dark current will greatly reduce the operation challenges.

The electron beam line illustrated in Fig.~\ref{fig:sketch}(b) was specially designed to demonstrate the feasibility of the DC-SRF-II gun as an electron source for CW XFELs. The Sol1 solenoid, whose center is at 1 m downstream the photocathode, is used for beam focusing and transverse emittance compensation. The emittance measurement device (EMD), based on a single slit scanning method~\cite{TAN2023167552}, comprises a motorized molybdenum plate with a 30~$\mu$m wide vertical slit (S1) for beamlet sampling and a YAG screen (Y3), integrated with a 45${}^\circ$ reflection mirror and a CCD camera, for electron divergence angle measurement. The EMD slit is located at 5.105 m downstream the photocathode, where the projected emittance at 100 pC bunch charge is expected to be well compensated according to simulation. It should be noted the electron beam from the gun has a rotationally symmetric distribution in the transverse directions, therefore only the horizontal EMD has been installed. The electron energy is measured with a dipole spectrometer, where a 90${}^\circ$ dipole magnet with a bending radius of R = 0.4 m is employed, {before which the beam is first focused by a solenoid lens (Sol2) and collimated by a 30~$\mu$m wide slit (S2) for a higher energy resolution.}

Compared to pulsed photocathode RF guns, the DC-SRF-II gun has a lower electric field at the photocathode surface. To mitigate the space charge induced emittance growth, the photocathode drive laser should have a longer pulse duration and a larger transverse size. In such a case, the electron beam would have a large transverse size before being focused by the Sol1 solenoid, as can be seen in Fig.~\ref{fig:sketch}(c). This makes the electron beam transport very sensitive to multipole field errors, especially of the emittance compensation solenoid where the electron beam size is a few millimeters (root mean square, rms). To mitigate the impact of the undesirable quadrupole and sextupole field components, two sets of coils are installed around the Sol1 solenoid, as shown in Fig.~\ref{fig:sketch}(b). The one before the solenoid, Qc, comprises eight coils configured to produce quadrupole field rotatable around the axis, while the other one, Sc, comprises six coils mounted on a rotating frame to produce rotatable sextupole field. 
{Another measure we have taken to mitigate the impact of multipole field errors is beam based alignment, which ensures the coincidence of the electron beam orbit with the axes of the SRF cavity and the Sol1 solenoid.}

The emittance optimization was focused on the high brightness operation mode at 100 pC bunch charge and 1 MHz rate. The DC voltage was at 100 kV and the cavity was operated {with an $E_\textrm{max}$ of about 22.3 MV/m}. The photocathode drive laser had a longitudinally quasi-plateau distribution with a length of 34 ps and a rising/falling edge of 6 ps, which was achieved through pulse stacking~\cite{Wang:24}. While in transverse plane, it had a truncated two-dimensional Gaussian distribution with the upper and lower limits at $\pm\sigma_0$, where $\sigma_0$, the standard deviation of Gaussian function, was 1 mm. {In this case, the drive laser has an rms size 0.48 mm on the photocathode.} Simulation studies show the optimal normalized emittance could reach 0.44 mm-mrad {with proper compensation under the above conditions (see Fig.~\ref{fig:sketch}) and the electron beam has a peak current of 6 A, a kinetic energy of 2.43 MeV, and a relative rms energy spread of 0.5\%}. Note the emittance could be further improved by increasing the cavity gradient or optimizing the temporal profile of the drive laser including reducing the rising/falling edge. 
{Also note that double emittance minimum effect~\cite{PhysRevE.55.7565} can be observed from the simulation result shown in Fig. 1(c), while the second emittance minimum could be shifted and get frozen at higher energy with proper matching of the beam into a downstream accelerator~\cite{PhysRevLett.99.234801}. 
In an earlier simulation study on a DC-SRF-II based 100 MeV injection line for CW XFELs, we have demonstrated a normalized emittance frozen below 0.4 mm-mrad and good longitudinal phase space performance under moderate operation conditions~\cite{ZHAO2021165796}.} 

\begin{figure}[htbp]
\begin{centering}
\vspace{-2mm}
\includegraphics[width=0.48\textwidth]{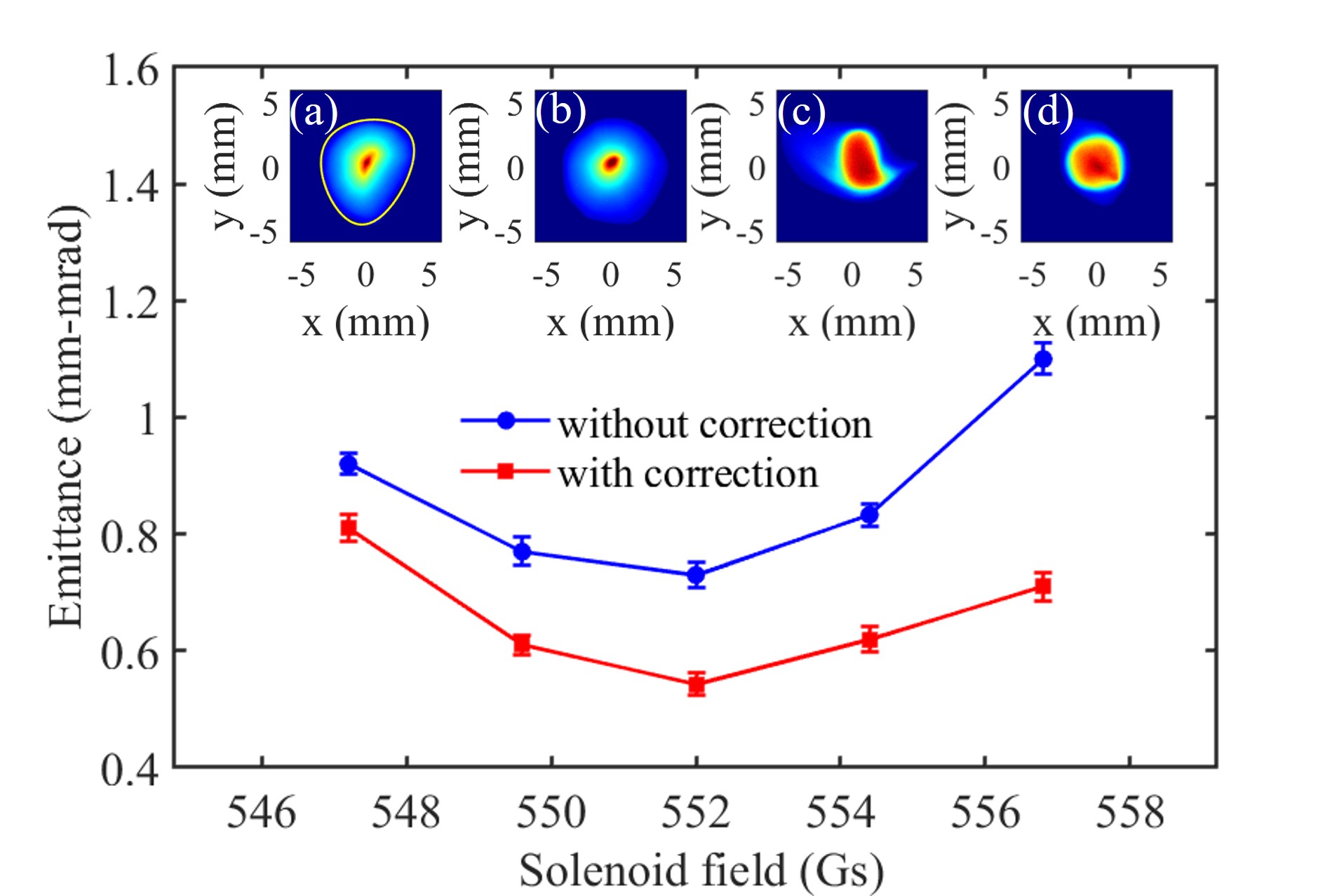}\\
\par\end{centering}
\vspace{-4mm}
\caption{\label{fig:emittance} {Measured normalized emittance} vs compensation solenoid strength before and after multipole field correction. Insets (a)/(c) show the transverse beam images at Y2/Y3 before corrections, while (b)/(d) show the images after corrections. }
\vspace{-4mm}
\end{figure}

To characterize the CW performance of the gun, a dedicated beam diagnostics mode was designed, for which the 1 MHz photocathode drive laser was modulated to generate low duty cycle electron bunch trains while keeping all other parameters the same as CW operation. 
The optimization process started with a two-dimensional scanning of the cavity phase and solenoid strength to determine the optimal phase for minimum emittance, which was $0^\circ$ (the on-crest acceleration phase) in the case reported herein. Then the acceleration phase was fixed at the optimal value and the electron beam phase space in the horizontal direction at different solenoid strengths was captured by the EMD with a small scanning step. To avoid the underestimation of emittance when excluding the invalid region of the captured images, two steps were taken to collect the data. First, the electron distributions along $x$ (position) and $x'$ (divergence angle) coordinates were both fitted to Gaussian functions with a standard deviation of $\sigma_x$ and $\sigma_{x'}$, respectively, and the phase space area within $(-3\sigma_x, 3\sigma_x )$ and $(-3\sigma_{x'}, 3\sigma_{x'})$ were extracted. Second, the data points for 5\% of the particles in the periphery region, which should contain some contribution from noise, was discarded. In this case, at least 95\% of the particles were included in the calculation. 
{A plot of the measured normalized emittance evaluated from the electron phase space} is shown in Fig.~\ref{fig:emittance}, which arrives at its minimum of 0.73 mm-mrad at the solenoid strength of 552 Gs (blue curve without correction yet). 

During the above experiments, a clear distortion of the electron distribution due to multipole magnetic field errors could be observed, as shown in Fig.~\ref{fig:emittance}. To quantify the effective multipole field components, the electron distribution was captured on the Y2 screen, where the electron beam still had a larger transverse size. The contour for the image was then extracted, as illustrated in Inset (a), from which the strength and angle of the effective quadrupole and sextupole fields were derived. For the optimal case presented in Fig.~\ref{fig:emittance}, the effective quadrupole component had an integrated field strength of 1.76 Gs and an orientation angle of $112^\circ$, while the effective sextupole component had an integrated field strength of 1.4 Gs/cm and an orientation angle of $31^\circ$. Subsequently, the quadrupole and sextupole correcting coils were set accordingly to cancel out the effect of multipole field errors. This led to a more regular and symmetric electron distribution shown in Insets (b) and (d). Such a correction was made at each instance in Fig.~\ref{fig:emittance} and the emittance was reduced significantly. Especially, for the optimal case, a normalized emittance of 0.54 mm-mrad has been achieved, which is close to the simulation result. 

\begin{figure}[htbp]
\begin{centering}
\vspace{-3mm}
\includegraphics[width=0.48\textwidth]{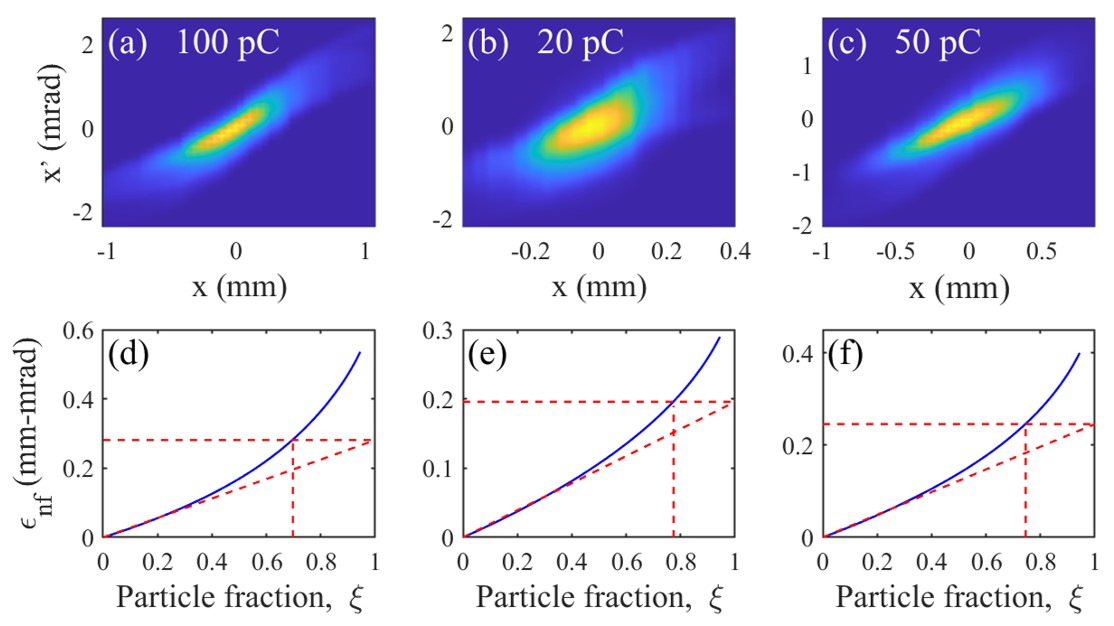}\\
\par\end{centering}
\vspace{-5mm}
\caption{\label{fig:phasespace}Electron beam phase space (a-c) and corresponding fractional normalized emittance vs particle fraction (d-f).}
\vspace{-3mm}
\end{figure}

Fig.~\ref{fig:phasespace}(a) shows the horizontal phase space of the electron beam with multipole field corrections, from which the fractional normalized emittance $\epsilon_{nf}$ as a function of particle fraction $\xi$ has been calculated, as plotted in Fig.~\ref{fig:phasespace}(d). The core emittance and core fraction, defined according to~\cite{PhysRevLett.102.104801}, have also been derived, which are 0.28 mm-mrad and 70\%, respectively, for this optimal case with 100 pC bunch charge. A summary of the parameters can be found in Tab.~\ref{tab:emittance}. 

The emittance optimization was also performed at 50 pC and 20 pC bunch charges with the same drive laser temporal profile as the 100 pC case. The phase spaces with multipole field corrections are shown in Fig.~\ref{fig:phasespace}, while the parameters are summarized in Tab.~\ref{tab:emittance}. {As a comparison, the NC RF gun for LCLS-II, the only CW XFEL facility in operation worldwide, is operated with an emittance of about 0.5 mm-mrad for 50 pC bunches~\cite{10.3389/fphy.2023.1150809}.}

\begin{table}[ht]
\centering
\vspace{-3mm}
\caption{\label{tab:emittance}Measured emittance and relevant parameters. {The drive laser 
has a same temporal profile for the three cases.}}
\vspace{-3mm}
\begin{tabular}[t]{lcccc}
\hline
Parameters & 100 pC & 50 pC & 20 pC & Units\\
\hline
{Cavity field ($E_\textrm{max}$)} & {22.3} & {21.7} & {21.7} & MV/m\\
Drive laser size {(rms)} & {0.48} & {0.48} & {0.38} & mm\\
Electron beam energy & 2.42 & 2.35 & 2.35 & MeV\\
Normalized emittance & 0.54 & 0.40 & 0.28 & mm-mrad\\
Core emittance & 0.28 & 0.25 & 0.19 & mm-mrad\\
Core fraction & 70\% & 75\% & 77\% & \\
\hline
\end{tabular}
\end{table}

\begin{figure}[htbp]
\begin{centering}
\vspace{-3mm}
\includegraphics[width=0.45\textwidth]{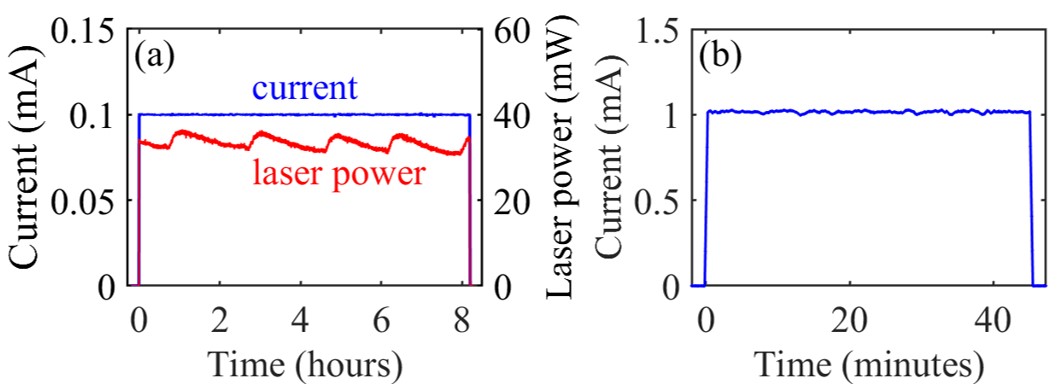}\\
\par\end{centering}
\vspace{-3mm}
\caption{\label{fig:cwoperation} Average beam current during CW operation tests at 1 MHz (a) and 81.25 MHz (b). The photocathode drive laser power is also plotted in (a).}
\vspace{-3mm}
\end{figure}

CW operations of the gun were mainly demonstrated at 1 MHz rate and 100 pC bunch charge. Fig.~\ref{fig:cwoperation}(a) shows the beam current monitoring results during a long-term run, which was recorded by the integrating current transformer (ICT) in Fig.~\ref{fig:sketch}(b). In the experiments, a beam current feedback based on photocathode drive laser attenuation was applied to maintain a constant current. The drive laser power is also plotted in Fig.~\ref{fig:cwoperation}(a), showing a cyclic change in a few hours' period. Such a phenomenon, only observed when the gun was operated in CW mode or high duty cycle quasi-CW mode, should be partially related to the local heating of the photocathode in its cryogenic environment, or the laser/mirror drifting at higher power. Although the mechanism for this variation needs to be further investigated, over one month’s operation with a single photocathode in the gun at a QE of about 1\% has already demonstrated the compatibility of the K${}_2$CsSb photocathode and the gun. 

Electron beam tests were carried out at 81.25 MHz rate, too. Fig.~\ref{fig:cwoperation}(b) shows the case at the kinetic energy of 1.7 MeV and average current of 1 mA. Short-term tests were also performed with a current up to 3 mA, while long-term operation at a higher current is expected in the future. 

{In conclusion, a high-brightness DC-SRF gun has been brought into stable CW operation. The performance parameters are on par with the NC RF guns for CW XFELs but with orders of magnitude lower dark current. In the operations, neither ion back-bombardment of photocathodes nor SRF cavity degradation due to photocathode contamination have been observed. These shows the  gun has become a competitive electron source candidate for CW XFELs and other MHz-rate beam applications. Further experiments will be carried out in the future to characterize the longitudinal phase of the electron beam.}

This work is supported by the National Key Re- search and Development Program of China (Grants No. 2016YFA0401904 {and  No. 2017YFA0701001}).

\bibliography{mybib}

\end{document}